\documentclass[conference]{IEEEtran}
\IEEEoverridecommandlockouts

\ifCLASSINFOpdf
   \usepackage[pdftex]{graphicx}
\else
\fi
\usepackage[cmex10]{amsmath}
\usepackage{array}

\begin{document}
\title{The Superconducting Toroid for the New International AXion Observatory (IAXO)}

\author{I. Shilon, A. Dudarev, H. Silva, U. Wagner and H. H. J. ten Kate

\thanks{Manuscript received July 13, 2013.}
\thanks{I. Shilon, A. Dudarev, H. Silva, U. Wagner and H. H. J. ten Kate are with the European Organization for Nuclear Research (CERN), CH-1211, Gen\`eve 23, Switzerland (phone: 0041-22-767-0963; e-mail: idan.shilon@cern.ch).}}

\maketitle

\begin{abstract}

IAXO, the new International AXion Observatory, will feature the most ambitious detector for solar axions to date. Axions are hypothetical particles which were postulated to solve one of the puzzles arising in the standard model of particle physics, namely the strong CP (Charge conjugation and Parity) problem. This detector aims at achieving a sensitivity to the coupling between axions and photons of one order of magnitude beyond the limits of the current detector, the CERN Axion Solar Telescope (CAST). IAXO is equivalent to combining roughly 20000 CAST detectors into a single apparatus. The IAXO detector relies on a high-magnetic field distributed over a very large volume to convert solar axions to detectable X-ray photons. Inspired by the ATLAS barrel and end-cap toroids, a large superconducting toroid is being designed. The toroid comprises eight, one meter wide and twenty one meters long racetrack coils. The assembled toroid is sized 5.2~m in diameter and 25~m in length and its mass is about 250 tons. The useful field in the bores is 2.5~T while the peak magnetic field in the windings is 5.4~T. At the operational current of 12~kA the stored energy is 500~MJ. The racetrack type of coils are wound with a reinforced Aluminum stabilized NbTi/Cu cable and are conduction cooled. The coils optimization is shortly described as well as new concepts for cryostat, cold mass, supporting structure and the sun tracking system. Materials selection and sizing, conductor, thermal loads, the cryogenics system and the electrical system are described. Lastly, quench simulations are reported to demonstrate the system's safe quench protection scheme.
\end{abstract}



%
\IEEEpeerreviewmaketitle

\section{Introduction}

The mathematical description of quantum chromodynamics includes a CP (Charge conjugation and Parity) symmetry violating term. However, experiments show that CP violation is not observed in strong interactions. This problem is commonly known as the strong CP problem. The most appealing solution to this problem extends the standard model by introducing an additional $U(1)$ symmetry, the so-called Peccei-Quinn (PQ) symmetry \cite{Peccei:1977hh, Peccei:1977ur}. The PQ symmetry breaks spontaneously to give rise to a new light neutral pseudoscalar particle, the axion, which is closely related to the neutral pion \cite{Weinberg:1977ma, Wilczek:1977pj}. Axions are currently one of the most interesting non-baryonic candidates for dark matter in the universe.

\begin{figure}[!h]
    \includegraphics[scale=0.5] {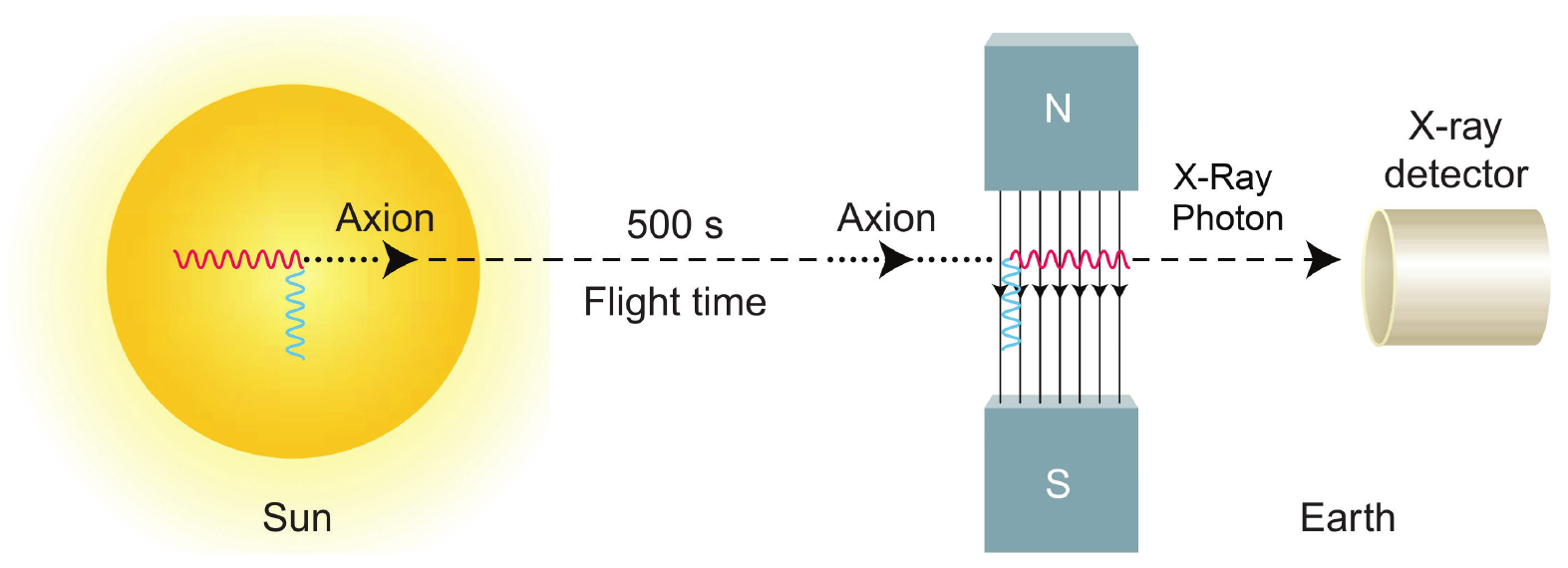}
    \caption{Schematic view of the solar axions detection concept: solar axions are predicted to convert to detectable X-ray photons when interacting with a magnetic field. Source: Science magazine Vol. 308 (2005).}
    \label{fig:1}
\end{figure}

Axions carry zero electromagnetic charge, while also being very light ($<10^{-36}$~kg) and spin-less. Hence, a direct observation of axions is practically impossible. However, axions are predicted to convert to and from photons when passing through an area of high magnetic field. Axions searches are exploiting this property by means of photons detection. In particular, an axion helioscope consists of a high field and large bore magnet, generating a magnetic field transverse to the solar axions flux, that is tracking the sun and equipped with X-ray detectors (see Fig. \ref{fig:1}). In this work we present the design of a new superconducting toroidal detector magnet, specifically designed and optimized for solar axions detection. This magnet constitutes the fundamental part of IAXO, the International AXion Observatory. The IAXO project entails a major upgrade of axion detection experiments, compared to the current state-of-the-art, which is represented by the CAST experiment at CERN and using a 9~T, 9~m long, LHC dipole prototype magnet with twin $15$~cm$^2$ bores. IAXO features a dramatic enhancement of the present limits on axions search.

\section{Principle Design}

\subsection{Figure of Merit and Lay-Out Optimization}

The driving design criterion of IAXO was defined as "achieving a sensitivity to the coupling between axions and photons of one order of magnitude beyond the limit of the current experiment (CAST)". To meet this criterion, the magnet's design is optimized according to its merit within the experiment. For axions detection, the magnet's figure of merit (MFOM) is defined by $f_M = L^2B^2A$ \cite{IAXO}, where $L$ is the magnet length, $B$ the effective magnetic field and $A$ the aperture covered by the X-ray optics. Currently, CAST's MFOM is 21~T$^2$m$^4$. The IAXO magnet has to attain an MFOM of 300 relative to CAST in order to meet the principal design criterion. Notice that the complete figure of merit of the experiment includes the tracking, detectors and optics parameters as well and is given in \cite{IAXO}.



\renewcommand{\arraystretch}{1.04}
\begin{table}[!ht]
\caption{Main Design Parameters of the IAXO Toroidal Magnet.}
\centering
\begin{tabular*}{0.45\textwidth}{@{\extracolsep{\fill}} p{5.55 cm}  c }
\hline\hline
\textit{Property} & \textit{Value}\\ 
\hline

\textbf{Cryostat dimensions:} \hfill Overall length (m) & 25  \\
\hfill Outer diameter (m) & 5.2 \\
\hfill Cryostat volume (m$^3$) & $\sim$ 530 \\

\textbf{Toroid size:} \hfill Inner radius, $R_{in}$ (m)  & 1.0 \\
\hfill Outer radius, $R_{out}$ (m) & 2.0 \\
\hfill Total axial length (m) & 21 \\

\textbf{Mass:} \hfill Conductor (tons) & 65 \\
\hfill Cold Mass (tons) & 130 \\
\hfill Cryostat (tons) & 35 \\
\hfill Total assembly (tons)  & $\sim$ 250 \\

\textbf{Coils:} \hfill Turns/coil & 180 \\
\hfill Nominal current, $I_{op}$ (kA) & 12.0 \\
\hfill Stored energy, $E$  (MJ) & 500 \\
\hfill Peak magnetic field, $B_p$ (T) & 5.4 \\
\hfill Inductance (H) & 6.9 \\
\hfill Average field in the bores (T) & 2.5  \\

\textbf{Conductor:} \hfill Overall size (mm$^2$) & 35 $\times$ 8 \\
\hfill Number of strands & 40 \\
\hfill Strand diameter (mm) & 1.3 \\
\hfill Critical current @ 5 T, $I_c$ (kA) & 58 \\
\hfill Operational margin & 40\% \\
\hfill Temperature margin @ 5 T  (K) & 1.9 \\

\textbf{Heat load:} \hfill  at 4.5 K (W) & $\sim$150 \\
\hfill at 60-80 K  (kW) & $\sim$1.6 \\
\hline\hline \end{tabular*} 
\label{table1}
\end{table}

To determine the MFOM, the magnet straight section length $L$ is set to 20~m and the integration $\int B^2(x,y)dxdy$ is performed over the \textit{open} area covered by the X-ray optics. Thus, the telescopes positioning must be defined to carry out the integration. Upon placing the telescopes as close as possible to the inner radius of the racetrack coil windings $R_{in}$, the optimized angular alignment of the telescopes is determined by the result of the integration. The latter indicates that the MFOM is affected mostly by the fraction of the telescopes aperture exposed to X-rays, implying that the telescopes optimized alignment is in between each pair of coils to allow for maximum telescopes exposure. The complete lay-out optimization study, including references to other magnet designs that were considered during the geometrical study, is described in detail in \cite{asc}.

\begin{figure}[!b]
\centering
    \includegraphics[scale=0.41] {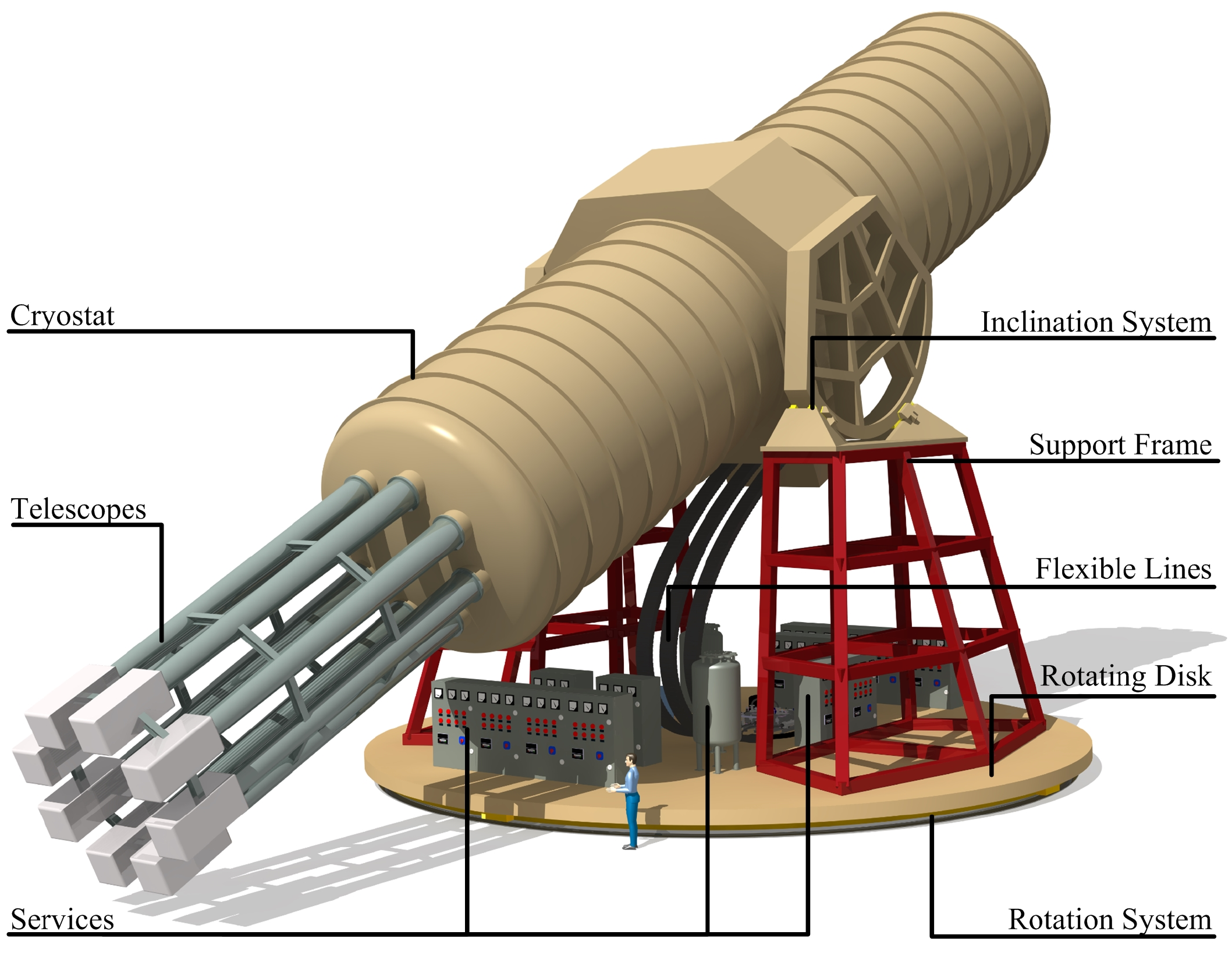}
    \caption{Schematic view of IAXO. Shown are the cryostat, eight telescopes, the flexible lines guiding services into the magnet, cryogenics and powering services units, inclination system and the rotating disk for horizontal movement. The dimensions of the system can be appreciated by a comparison to the human figure positioned near to the rotating table.}
    \label{fig:2}
\end{figure}

The magnet system design, presented in Fig. \ref{fig:2}, follows the results of the geometrical optimization study. The design meets all the experimental requirements of the magnet. It is relying on known and mostly well-proven engineering solutions, many of which were used in and developed for the ATLAS toroids, thus making the magnet system technically feasible to manufacture. The main properties of the toroid are listed in Table 1. The design essentially features a separation of the magnet system from the optical detection systems, which considerably simplifies the overall system integration. This allows for eight open warm bores in between the racetrack coils, as follows from the lay-out optimization study. The inclusion of the warm bores will simplify the fluent use of experimental instrumentation and the periodic maintenance~of~the system. 

The toroidal magnet comprises eight coils and their casing, an inner cylindrical support for the magnetic forces, keystone elements to support gravitational and magnetic loads, a thermal shield, a vacuum vessel and a movement system (see Figs. \ref{fig:2} to \ref{cs} and Table 1). Its mass is $\sim$250~tons. At the operational current of 12~kA the stored energy is $\sim$500~MJ. The design criteria for the structural design study are: a maximum deflection of 5 mm, a general stress limit of 50~MPa and a buckling factor of 5. The magnetic and structural designs are performed using the Maxwell 16.0 code and an ANSYS$^{\textregistered}$ 14.5 Workbench environment. The eight bores are facing eight X-ray telescopes with a diameter of 600~mm and a focal length of $\sim$6~m. The diameter of the bores fits the diameter of the telescopes. The choice for an eight coils toroid with the given dimensions and eight 600~mm diameter telescopes and bores is also determined by a cost optimization within the anticipated budget for construction of the magnet of about 30 MEuro.  


\subsection{Conductor}

\begin{figure}
    \includegraphics[scale=0.27] {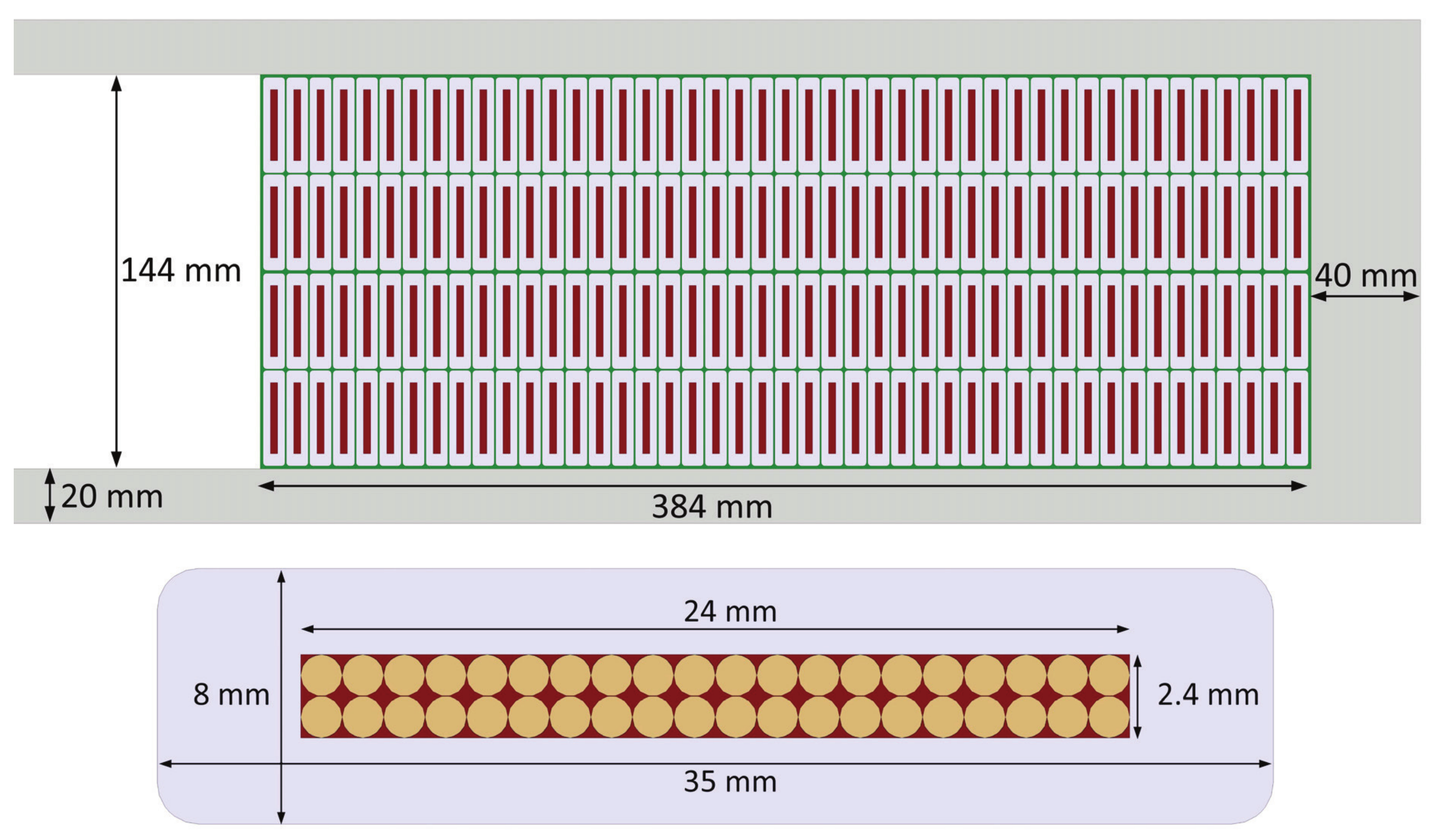}
    \caption{Cross section of the two double pancake winding packs, the coil casing (top) and the conductor with a 40 strands NbTi/Cu Rutherford cable embedded in a dilute Al-0.1wt\%Ni doped stabilizer (bottom).}
    \label{coil}
\end{figure}

The conductor is shown in Fig. \ref{coil}. The Rutherford type NbTi/Cu cable, composed of 40 strands of 1.3~mm diameter and a Cu/NbTi ratio of 1.1, is co-extruded with a reinforced high RRR aluminum stabilizer, following the techniques used in the ATLAS and CMS detector magnets at CERN \cite{ATLAS}-\cite{CMS}. The critical current of the conductor is $I_c(5~\mbox{T},~4.5~\mbox{K})~=$~58~kA.

The peak magnetic field in the windings is 5.4~T for a current of 12~kA per turn. The net force acting on each racetrack coil is 19~MN, directed radially inwards. The racetrack coils are bent to a symmetric arc shape, with radius $R_{arc} = 0.5$~m, so to minimize the forces acting on the bended sections.

To maximize its MFOM, the IAXO toroid needs to maintain the highest possible magnetic field. Nonetheless, adequate operational current and temperature margins are obligatory in order to ensure proper and safe operation. For a two double pancake configuration with 180 turns and engineering current density $J_{eng} = 40$~A/mm$^2$, the peak magnetic field is $B_p = 5.4$~T. The critical magnetic field corresponding to the magnet load line is 8.8~T at 65~A/mm$^2$. Hence, IAXO's magnet is expected to work at about 60\% on the load line, setting the operational current margin to 40\%.

The temperature margin calculation is based on an operational temperature of $T_{op} = 4.5$~K and a peak magnetic field of $B_p = 5.4$~T. A coil with two double pancakes and 45 turns per pancake satisfies a temperature margin of 1.9~K.

\subsection{Electrical Circuit and Quench Protection}

The adiabatic temperature rise in the case of a uniformly distributed quench is $\sim$ 100~K. The toroid's quench protection is based on an active system and an internal dump of the stored energy. The principle of the quench protection system is to rely on a simple, robust and straight forward detection circuits, usage of simple electronics and at least a three fold redundancy in order to reduce failure probability.

\begin{figure}[b]
\centering
    \includegraphics[scale=0.49] {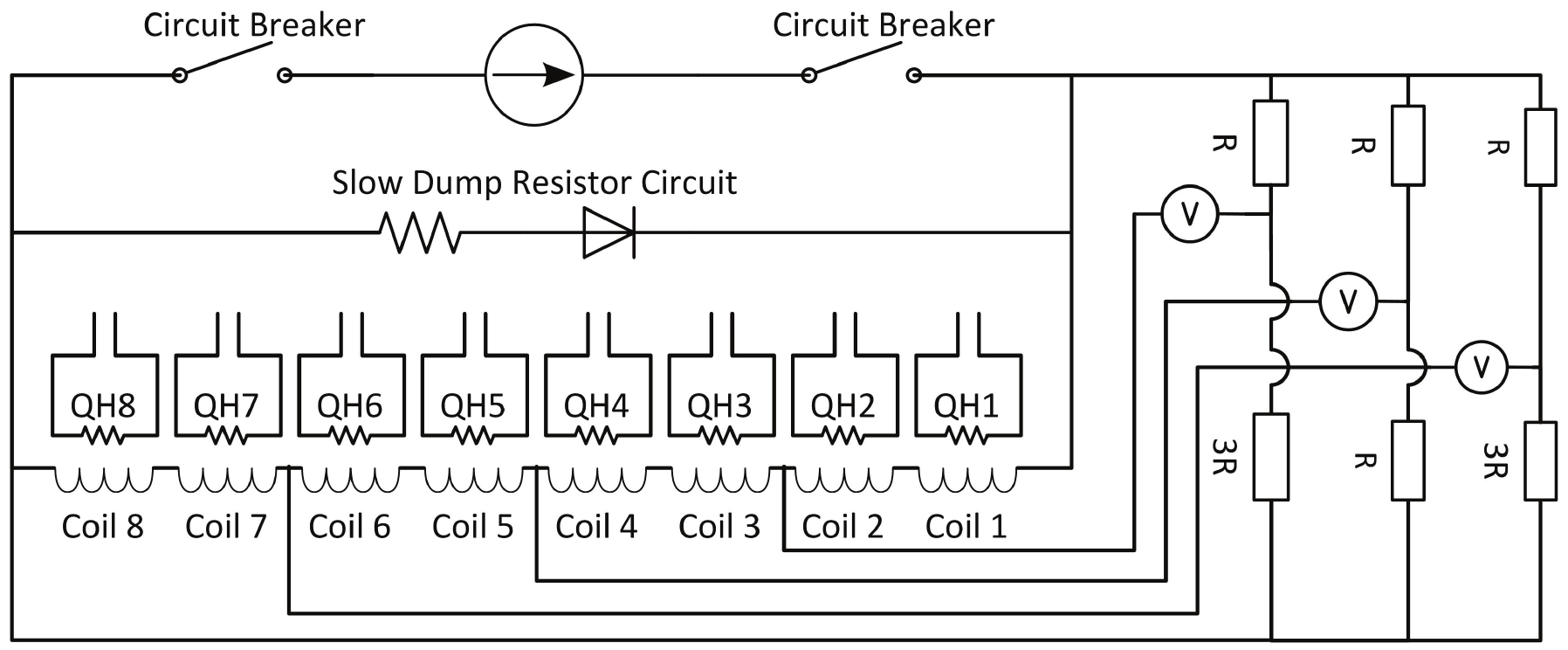}
    \caption{Schematic diagram of the electrical circuit and quench protection scheme. Shown are the power convertor, the eight coils, quench heaters (QH 1-8), the slow dump circuit and the quench detection circuit.}
    \label{el}
\end{figure}

The electrical circuit of the IAXO toroid is shown in Fig. \ref{el}. The magnet power convertor of 12~kA is connected at its DC outputs to two breakers, which open both electrical lines to the magnet. The high-$T_c$ current leads are installed within their own cryostat, that is integrated on the rotating gantry of the magnet. The current leads feed the eight coils, which are connected in series, by means of flexible superconducting cables. The flexibility is required to compensate for the changing inclination of the racetrack coils. Each coil is equipped with multiple quench heaters, connected in parallel. Across the warm terminals of the current leads, a slow-dump-resistor with low resistance is connected in series to a diode. The quench detection circuit relies on the detection of normal voltage growth across the toroid following a quench. The voltage sensitivity level of the detectors is 0.3~V, which implies a typical detection delay time of $\sim$ 1~s. The protection circuit is equipped with a timer delay so that false signals do not activate the protection system and lead to a bogus fast discharge. 

When a quench is detected and verified, the two breakers open to quickly separate the magnet from the power convertor and a quench is initiated in all coils simultaneously by activating all the quench heaters. This ensures a fast and uniform quench propagation and thus a homogenous cold mass temperature after a quench. Simultaneously, the current is discharged through the dump resistor. This discharge mode, the so-called fast dump mode, is characterized by an internal dump of the magnet's stored energy. Upon choosing a coil symmetric grounding point, the fast discharge scheme ensures that the discharge voltage excitation is kept low enough and that the stored energy is uniformly dissipated in the windings. The internal energy dump depends on the absolute reliability of the quench heaters system. To reduce failure probability to an acceptable level all components feature a three fold redundancy.

The DC power convertor will operate in voltage control mode when ramping up the toroid and in current control mode during steady operation. The field stability requirement for an axion helioscope is of minor importance. A time stability as large as 0.1\% will not affect for the axion-photon conversion probability, and hence in the experiment's sensitivity. 

Under normal operation, the toroid will be discharged through the diode-resistor circuit in a passive run down mode (slow discharge mode). Slow discharge is also the safety dump mode activated in the case of a minor fault. 

Each of the dump-resistors is connected in series to a diode unit to avoid current driven through the dump resistor circuit during normal operation of the magnet. The dump resistors circuit is air cooled by convection and have the capacity of absorbing the total stored magnetic~energy~of~the~toroid.

The longitudinal normal zone propagation velocity is 6~m/s. The velocity was calculated by using COMSOL 4.3b coupled multiphysics modules~in~a~2D adiabatic model. Hence, the normal zone will propagate around an entire coil (43~m) in 3.5~s.

\subsection{Cold Mass}

\begin{figure}[!b]
\begin{center}
    \includegraphics[scale=0.397] {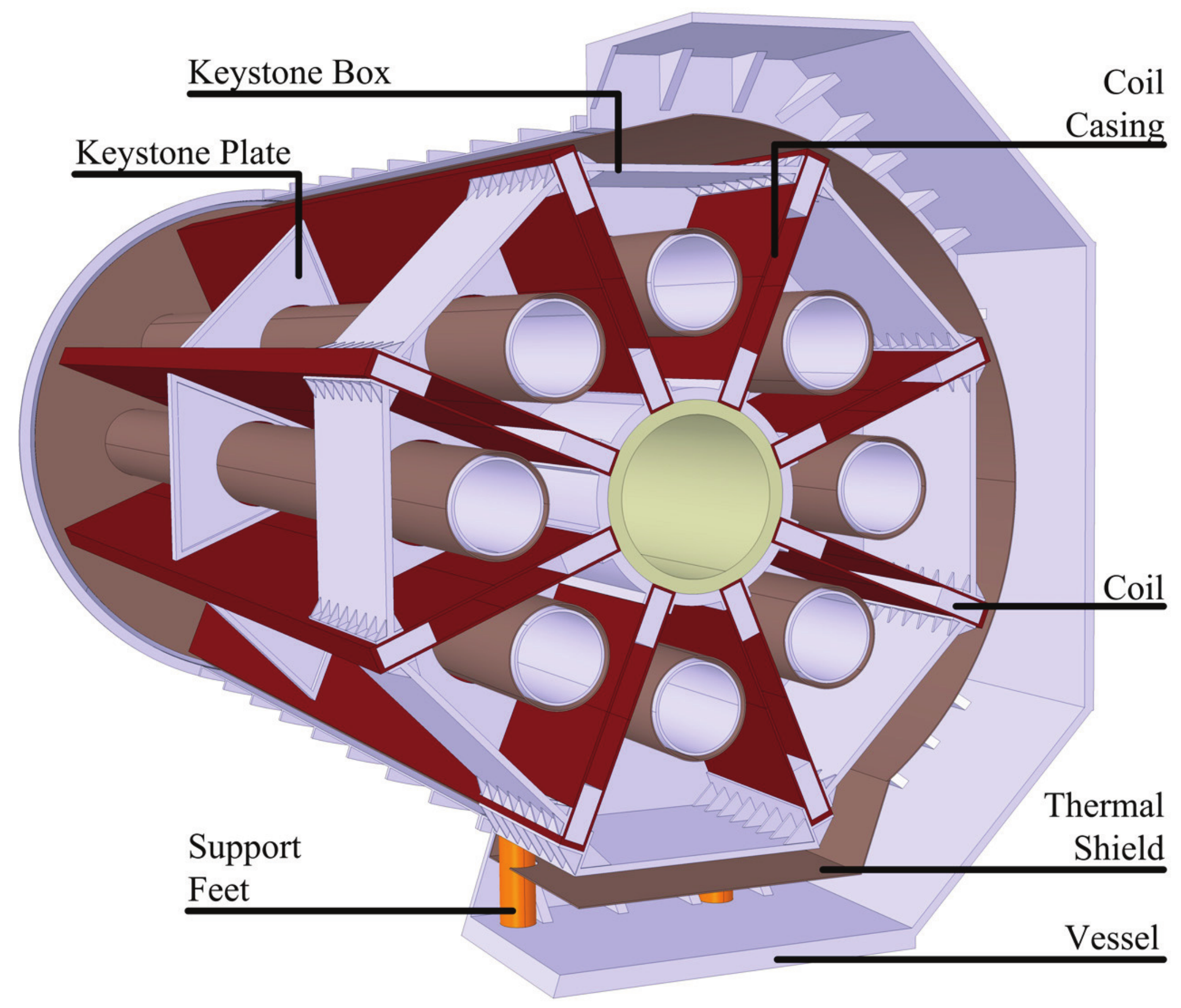}
    \caption{Mid-plane cut of the cryostat with an exposed cold mass, showing the cold mass and its supports, surrounded by a thermal shield, and the vacuum vessel. The open bores will allow for easy periodic maintenance of the magnet system while simplifying the use of experimental instrumentation.}
    \label{cs}
\end{center}
\end{figure}

The cold mass operating temperature is 4.5~K and its mass is approximately 130~tons. It comprises eight coils, with two double pancakes per coil, which form the toroid geometry, and a central cylinder designed to support the magnetic force load. The coils are embedded in Al5083 alloy casings, that are attached to the support cylinder at their inner edge. The casings are designed to minimize coil deflection due to the magnetic forces. To increase the stiffness of the cold mass structure and react gravitational and magnetic loads, and to support the warm bores, eight Al5083 keystone boxes and 16 keystone plates are connected in between each pair of coils, as shown in Fig. \ref{cs}. The keystone boxes are attached to the support cylinder at the center of mass of the \textit{whole} system (that is, including the telescopes and detectors) and the keystone plates are attached at half-length between the keystone boxes and the coils ends. 

A coil, shown in Fig. \ref{coil}, consists of two double pancake windings separated by a 1 mm layer of insulation. The coils are impregnated for proper bonding and pre-stressed within their individual casing to minimize shear stress and prevent cracks and gaps appearing due to thermal shrinkage on cool down and magnetic forces.

\subsection{Cryostat and Its Movement System}

The cryostat comprises a rigid central part, placed at the center of mass of the \textit{whole} system, which also serves as a fixed support point for the cold mass, with two large cylinders and two end plates enclosing it to seal the vacuum vessel. Eight cylindrical open bores are placed in between the end plates. The vessel is optimized to sustain the atmospheric pressure difference and the gravitational load, while being as light as possible. The Al5083 rigid central post is 70~mm thick with a 150~mm bottom plate to support the weight of the cold mass. Using two end flanges at the vessel's rims, as well as periodic reinforcement ribs at 1.35~m intervals along both cylinders, the structural requirements are met for a 20~mm thick Al5083 vessel with two 30~mm thick torispherical Kl\"opper shaped end plates. The 10~mm wall thickness of the eight cylindrical bores is minimized for the bores to be placed as close as possible to the racetracks coils inner radius. The cold mass supports are made of four G10 feet, connecting the reinforced bottom keystone box to the central part of the cryostat and by that transferring the weight load of the cold mass to the cryostat. 

The IAXO detectors need to track the sun for the longest possible period. The vertical inclination range is $\pm$ 25$^\circ$, while the horizontal rotation is a full 360$^\circ$, after which the magnet revolves back to its starting position. The 250 tons magnet system is supported at the center of mass of the whole system through the cryostat central part (see Fig. \ref{fig:2}), thus minimizing the torques acting on the support structure. An altitude-over-azimuth mount configuration, commonly used for very large telescopes, was chosen to support and rotate the magnet system. The vertical movement is performed by two semi-circular structures (C-rings) with extension sections attached to the central part of the vacuum vessel. The C-rings are mounted on top of a 6.5~m high structural-steel support frame, situated on a wide rotating structural-steel disk. The rotation of the disk is generated by a set of roller drives on a circular rail.

The required magnet services, providing vacuum, helium supply, current and controls, are placed on top of the disk to rotate along with the magnet. The magnet services are connected via a turret aligned with the rotation axis, thus simplifying the flexible cables and transfer lines arrangement. Flexible chains guide the services lines and cables from the different services boxes to the stationary connection point.

\subsection{Cryogenics}

The heat load on the magnet by radiation and conduction is $\sim$150~W at 4.5~K. The thermal shield heat load is $\sim$1.6~kW at 60-80~K. An acceptable thermal design goal is to limit the temperature rise in the coils to 0.1~K above the coolant temperature under the given heat loads. The coil windings are conduction cooled at a temperature of 4.5~K. The conceptual design of the cryogenic system is based on forced flow cooling of sub-cooled liquid helium at a supercritical pressure. This avoids two-phase flow within the magnet cryostat and hence the complexity of controlling such a flow within a system whose inclination angle is continuously changing. The coolant flows in a piping system attached to the coil casings, allowing for conduction cooling in a manner similar to the ATLAS toroids \cite{ATLAS, ATLAS2}. The cryogenic system concept features a helium compression and gas management that are ground stationary. A refrigerator cold box, current leads cryostat and a 4.5~K liquid helium buffer and valve box are integrated on the rotating disk that carries the structure of the helioscope. A helium bath is connected to the magnet~cryostat~to follow~its~movement. The cryogenic system is described in greater detail in \cite{icmc}.

\section{Conclusion}

The design of the new IAXO toroidal superconducting magnet satisfies the design criterion of increasing the solar axion searches sensitivity by at least one order of magnitude. The design relies on known engineering solutions and manufacturing techniques and thus is technically feasible to manufacture. With the inclusion of eight open warm bores, the design also allows maximum flexibility for experimentalists. The magnet system and optical detectors are separated, thus allowing for a parallel effort of development, construction and initial operation, thereby also minimizing cost and risks. In the coming years the design will be further detailed and project funding will be proposed.






%

\end{document}